\documentstyle[prl,aps,epsfig,twocolumn]{revtex}
\begin{document}
\twocolumn[\hsize\textwidth\columnwidth\hsize\csname@twocolumnfalse\endcsname
\title{Experimental evidence of dynamical localization and
  delocalization in a quasi-periodic driven system}
\author{J. Ringot, P. Szriftgiser, and J. C. Garreau}
\address{Laboratoire de Physique des Lasers, Atomes et Mol{\'e}cules and
Centre d'Etudes et de Recherches Laser et Applications,
Universit\'{e} des Sciences et Technologies de Lille, F-59655 Villeneuve d'Ascq Cedex, France\\}
\author{D. Delande}
\address{Laboratoire Kastler-Brossel,
Tour 12, Etage 1, Universit\'e Pierre et Marie Curie, 4 Place Jussieu, F-75005 Paris} 
\maketitle

\begin{abstract}
This paper presents the first experimental evidence of the transition
from dynamical localization to delocalization under the influence
of a quasi-periodic driving on a quantum system. A quantum kicked
rotator is realized by placing cold atoms in a pulsed, far-detuned, 
standing wave. If the standing wave is periodically pulsed,
one observes the suppression of the classical chaotic diffusion,
i.e. dynamical
localization. If the standing wave is pulsed quasi-periodically
with two different frequencies, 
dynamical localization is observed or not, depending on the two 
frequencies being commensurable or incommensurable. One can thus
study the transition from the localized to the delocalized case
as a function of the effective dimensionality of the system.
\end{abstract}


]

Dynamical localization (DL) is a 
specifically quantum phenomenon taking place in
time-periodic systems whose corresponding classical
dynamics displays chaotic diffusion. While, in the classical
limit, because of the diffusion process, the system
spreads indefinitely in the phase space, the quantum system 
follows the classical diffusive
dynamics for short time only, but after some
{\em localization time} freezes its evolution with
no further increase of the average energy.

This behavior, attributed to quantum interferences among 
the diffusive paths which for long times are in the average
completely destructive, was numerically
observed at the end of the 70's on the one-dimensional kicked rotator
exposed to periodic kicks \cite{ref:DL_First}, a paradigmatic 
simple system whose classical dynamics can be reduced to iterations of the 
Chirikov's standard map.

The possibility of observing DL with a system
constituted of cold atoms placed in a far-detuned
standing wave has been theoretically suggested in 1992
\cite{ref:Graham} 
and experimentally observed in 1994 \cite{ref:Raizen}.
A crucial question is whether DL is robust versus
perturbations of the system. Indeed, as it strongly relies on
quantum interferences, it is expected to be rather fragile.
As a matter of fact, it has been experimentally shown that DL 
can be partly or totally destroyed by decoherence (i.e. coupling of
the system to external degrees of freedom; in the present context 
spontaneous emission plays such a role) and noise, that is 
deviation from strict periodicity \cite{ref:DLDissip,ref:DLNoise}. 

Moreover, there is a relevant
connection of DL with the Anderson localization taking place
in disordered systems. Indeed, the periodically kicked rotator
problem can be mapped on a one-dimensional Anderson model, that is a model
of a particle moving along a one-dimensional chain of sites \cite{fishman},
with coupling between neighbors and diagonal disorder, i.e. 
pseudo-random values of the potential energies on each site.
In 1-d, although the classical motion is diffusive, the
quantum eigenstates are all localized. Because of the similarity
of the Anderson and kicked top models, Anderson localization and DL
are similar quantum phenomena, the first one taking place along
space coordinates, the second one along the time coordinate with 
localization in the momentum domain.
It is well known that Anderson localization is strongly
dependent on the number of spatial degrees of freedom.
Similarly, DL is expected to be highly sensitive on the number
of temporal degrees of freedom, that is on the frequency
spectrum of the external driving. 

We here consider the interesting simple case where the
external driving is not periodic but quasi-periodic, with
two independent frequencies. Theoretical arguments and numerical
simulations \cite{ref:Multidim} suggest that the situation is similar
to the 2-d Anderson model, and that the localization time should 
become so large
that it might be impossible to observe DL experimentally. 
The goal of this paper is to study {\em experimentally}
such a situation of increased dimensionality 
and to test the theoretical predictions.

We realized a quantum kicked rotator with a primary series of
kicks of frequency $f_1$ to which a secondary
series of kicks (frequency $f_2$) can be added, with
$f_2/f_1 = r$.
A physical experiment cannot be sensitive to the rational
or irrational character of a number; one might thus consider
only rational values of this frequency ratio. 
For a given rational value of $r=p/q$ (irreducible fraction), 
the periodicity of the system 
cannot have any physical effect before at least $q$ kicks
of the primary series. 
Thus, periodicity effects like DL cannot show up unless the
number of primary kicks is large compared to $q$. As the number of
primary kicks increases, one expects to find more and more
``rational" values of $p/q$ for
which DL is effectively observed. Working with a
two-frequency quantum rotator thus allows one to go
from the 1-d to the 2-d case by choosing irreducible frequency 
ratios corresponding to larger and larger $q$ \cite{ref:Multidim}.
Experimental results on the two-frequency microwave ionization
of Rydberg atoms have indirectly shown the importance of 
rational  or irrational values in quantum transport
properties \cite{ref:Koch}.

The atomic quantum kicked rotator is realized by placing cold 
atoms (cesium in our case) of mass $M$ in a far-detuned, 
pulsed standing wave of intensity $I_0$, wave
number $k_L$ and detuning $\Delta$ with respect to the closest
atomic transition (the cesium $D2$ line at 852 nm). If the detuning is large
enough, the dominant interaction between atoms and the laser light is
the light-potential which is proportional to the intensity. One
then obtains a Hamiltonian of the form
\begin{equation}
H = { p^2 \over 2M } - V_0 \cos(2k_Lx) f(t)
\label{eq:Hamiltonian}
\end{equation}
where $f(t)$ is a function of period $T$,
and $V_0 = \hbar \Omega^2/8\Delta,$ 
where $\Omega$ is the resonant
Rabi frequency. $V_0$ is proportional to the light intensity. 

In the limit where the width of the peaks in 
$f(t)$ is negligible compared to $T$ (i.e., each peak approaches 
a delta function), rescaling variables
\cite{ref:Graham} allows one to reduce this Hamiltonian
to the standard form corresponding to the quantum rotator: 
\begin{equation}
{\cal H}_1 = { P^2 \over 2 } - K \cos\theta \ \sum_n{\delta(\tau-n)}
\label{eq:H}
\end{equation}
where $K$ is the so-called {\em stochasticity parameter} and
where the new conjugate variables obey the quantum commutation
rule $[\theta,P] =i \hbar_{\mathrm eff}$
with $\hbar_{\mathrm eff}=4 \hbar k_LT/M$ the {\em effective Planck
constant}. The classical limit ($\hbar_{\mathrm eff} \rightarrow 0$) 
of such a system becomes (weakly) chaotic for 
$K \approx 1$ and fully chaotic for $K \approx 10$.
When a second series of pulses is applied, the reduced Hamiltonian becomes:
\begin{eqnarray}
{\cal H}_2 = { P^2 \over 2 } - \cos\theta  \{ K_1
\sum_n{\delta(\tau-n)} + \nonumber \\
K_2 \sum_n{\delta [\tau-(n+\phi/2 \pi)/r ] }  \} 
\label{eq:H2}
\end{eqnarray}
with $K_1=K_2$ in our experiment.
In the above equation $\phi$ is the phase of the second series of pulses
with respect to the first series.
The classical dynamics of this system is essentially identical to the
periodic kicked rotator: for $K_1=K_2 \approx 10,$ it is a chaotic diffusion.

\begin{figure}
\begin{center}
  \epsfig{figure=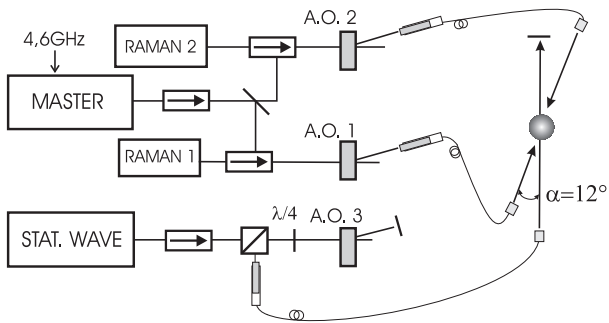,width=8cm,clip=}
\end{center}
\caption{Experimental setup. A master diode laser modulated
at 4.6 GHz is used to inject two power slave Raman lasers
producing phase-coherent, 9.2 GHz frequency-split, Raman
beams. A power monomode diode laser is used to generate
the stationary wave, that can be pulsed through an acousto-optical
modulator (mounted in double passing).
Both the Raman and stationary wave beams are horizontal,
making an angle of $12^\circ$.}
\label{fig:setup}
\end{figure}

Our realization of the kicked rotator (Fig. \ref{fig:setup})
is similar to that of ref.
\cite{ref:DLNoise}. Cold cesium atoms issued from a magneto-optical
trap (MOT) are placed in a 
far-detuned, pulsed standing wave. The measurement of 
momentum distribution is accomplished in our setup by
velocity-selective
Raman stimulated transitions
between the $F_g=3$ and $F_g=4$ hyperfine ground-state sublevels
\cite {ref:SBCooling}.
Generation of the Raman beams is based on direct current modulation
at 4.6 GHz of a diode laser, detuned by 200 GHz with respect 
to the atomic transition. The two symmetric first-order optical 
sidebands are then used to inject two 
diode lasers that produce 150 mW beams with a 9.2 GHz beat-note
of sub-hertz spectral width \cite{ref:DiodeModulee}.

Cesium atoms are first 
optically pumped into the $F_g=3$ hyperfine sublevel.
A Raman pulse of detuning
$\delta_R$ brings the atoms in the velocity class $v=\delta_R/(2k_R)$ 
($k_R$ is the wave number of the Raman beams) back to the 
$F_g=4$ hyperfine sublevel. A probe beam resonant with the
transition from the sublevel $F_g=4$ is frequency modulated, and 
its absorption signal detected by a lock-in amplifier, yielding a signal
proportional to the population of the $F_g=4$ level.

Stray magnetic fields are 
harmful for the Raman velocity measurement.
3D-magneto-resistive probes are placed at the eight
corners of the MOT cell. Their signal is electronically interpolated and
generates a feedback signal to
three mutually orthogonal Helmholtz coil pairs \cite{ref:ArticleLong}.
We measured a residual magnetic field
below 250 $\mu$G and an effective compensation band-width of 500 Hz. 
The $\hbar k_L/2$ momentum resolution then obtained is largely
sufficient for this experiment, and is much better than that obtained
by time of flight methods \cite{ref:DLNoise}.

A power diode laser is detuned by
7 GHz with respect to the cesium D2 line at 852 nm.
An acousto-optical modulator is used to generate
arbitrary series of pulses.
The modulated beam is then
transported by an optical fiber to the neighborhood of 
the MOT apparatus. The standing wave, obtained by 
back-reflection of this beam, has a waist
of 0.6 mm and a typical power of 50 mW in each direction. It is 
modulated with two series of pulses: the
primary pulses of fixed frequency $f_1=36$ kHz are 500 ns long, corresponding
to a stochasticity parameter $K_1=10$
and to an effective Planck constant
$\hbar_{\mathrm eff} = 2.9.$ The pulse shape
is rectangular with a 
rise and fall time of the order of 50 ns.
The secondary pulses have the same duration and
the same intensity, but their frequency $f_2$ and phase $\phi$ can be 
adjusted at will. A typical experiment is done with 50 primary pulses. 
In order to avoid pulse superposition effects between the two series,
the phase $\phi$ is fixed to an arbitrary non-zero value.

\begin{figure}
\begin{center}
  \epsfig{figure=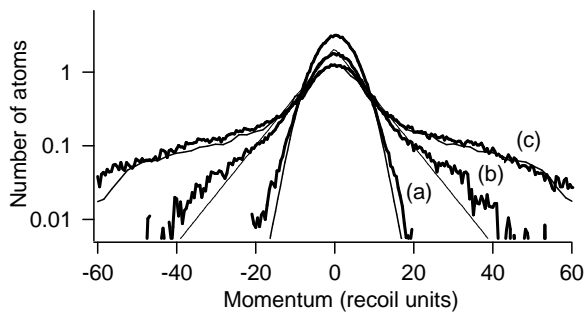,width=8cm,clip=}
\end{center}
\caption{Typical momentum distributions (in logarithmic scale)
corresponding to:
(a) initial distribution produced by the MOT, corresponding
to a temperature of 3.3 $\mu$K; the fitting curve (thin line) is
a gaussian. (b) Momentum distribution
obtained after the interaction of the atoms with two series of
kicks having $f_2=f_1=36$ kHz and a relative phase $\phi=180 ^\circ$;
it displays the exponential shape characteristic of dynamical localization
for a time-periodic quantum system; the fitting curve is exponential
(thin line).
(c) Momentum distribution
after interaction with two series of kicks having $f_2/f_1=1.0833$
and initial relative phase $\phi=180 ^\circ$. The distribution is broader,
indicating the destruction of dynamical localization in a quasi-periodic
driven quantum system; the fitting curve is a numerical
simulation (thin line); for details, see text.
The recoil momentum is $\hbar k_L$.}
\label{fig:VelDistr}
\end{figure}

In an experimental run, cesium atoms are first
cooled and trapped by the MOT.
A Sisyphus-molasses phase further
reduces the temperature to about 3.3 $\mu$K.
The MOT beams are turned off and a pulse of a
repumper beam transfers the atoms from the $F_g=4$ to the 
$F_g=3$ hyperfine sublevel. The standing wave is then 
turned on. When the standing wave excitation ends, the Raman
sequence described above is used to detect the population of
a velocity class. The whole sequence then starts over with a
different value of the Raman detuning to probe a new
velocity class.
The pulse sequence is produced by two synthesizers
at frequencies $f_1$ and $f_2$ with a fixed phase relation.
We show in Fig. \ref{fig:VelDistr} the initial 
momentum distribution (just before the kicks are applied)
and the final distributions (after interaction with
the standing wave) for $f_2/f_1=1.000$ and $f_2/f_1 = 1.083$
and a phase of $\phi = 180^\circ$.
The initial distribution is a gaussian with a typical full width
at half maximum (FWHM) of $10 \hbar k_L$. 
Both final distributions show a clear 
broadening with respect to the initial one. 
For the ``resonant" case
($f_2/f_1=1$) [trace (b)], the distribution presents
a characteristic exponential
shape $P(p) \simeq \exp{(-|p|/L)}$, with a
localization length (along the momentum axis) 
$L \approx 8.5 \hbar k_L$,
which is a signature of the dynamical localization.
This is not surprising as for $f_1=f_2,$ the system is strictly
time-periodic and thus should present dynamical localization.
The measured localization length agrees fairly well with 
theoretical estimates.
Trace (c) corresponds to a non-resonant truly quasi-periodic case,
where the ratio 
$f_2/f_1=1.083$ is sufficiently far from any simple rational
number. The momentum distribution presents a broader and
more complex shape.
We have performed numerical simulations of the system, as described by 
Eq.~(\ref{eq:H2}): we have solved ``exactly" the Schr\"odinger equation
using a method similar to the one described in \cite{cohen}. The resulting
momentum distribution is averaged over the measured initial momentum 
distribution of the atoms
and over the inhomogeneous laser intensity. We have used $K_1=K_2=10$
at the center of the laser beam, in accordance with the value deduced
from the laser power, detuning and geometrical properties.
The only adjustable parameter is the ratio of
{\em effective} sizes of the standing wave and the Raman
beams. Due to the nonlinearity of the processes, this ratio
(which is 2) is different from ratio of the waists (4.8).
For $f_1=f_2=36$ kHz, we obtain a dynamically
localized (exponential) distribution with a localization length
which agrees with the experimentally observed one (at the 10\% level).
For $f_2/f_1=1.083,$ the result of the simulation -- shown in the figure --
agrees very well with the experimental data.

\begin{figure}
\begin{center}
  \epsfig{figure=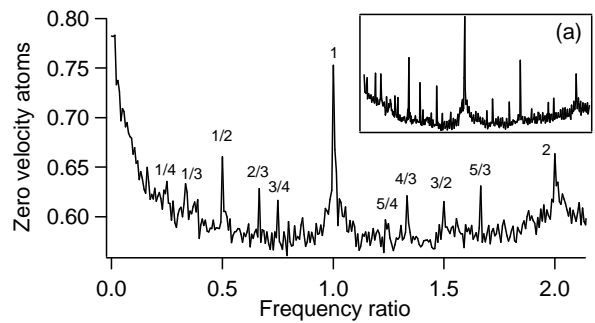,width=8cm,clip=}
\end{center}
\caption{The population of zero-velocity atoms (probed
with the Raman signal) as a function of the frequency
ratio $r=f_2/f_1$ (with $ f_1=36$ kHz) and phase
$\phi = 52 ^\circ$.
The increase of the zero-velocity signal is a signature of
dynamical localization. Dynamical localization for
commensurate frequencies -- and simple rational $r$ values -- is clearly seen.
For incommensurate frequencies, like in Fig.~\protect{\ref{fig:VelDistr}},
no dynamical localization is visible. The inset (a) shows the
corresponding curve obtained by numerical
simulation (see text), very well reproducing the features
of the experimental curve.}
\label{fig:LocSpectrum}
\end{figure}

The fact that the broad contribution is significantly larger
than the ``resonant" distribution -- together with
the fact that the classical diffusion constant is practically
identical in the two cases -- shows that diffusion has persisted
during a longer time in the non-resonant case. Furthermore, the
fact that the distribution is not exponential strongly suggests
that we did not reach DL and that diffusion should persist for longer times. 
A simple and useful method to detect the presence of DL
is to probe only the zero-velocity class: as DL
corresponds to a thiner distribution, it also corresponds to
a higher zero-velocity
signal in the localized case than in the diffusive case. In other
words, the zero-velocity signal contains essentially the
same information than e.g. the total average energy, but is much
easier to measure. This allows
us to sweep the frequency $f_2$ of the secondary kick, keeping 
all other parameters ($f_1$, $\phi$, $K_1$ and $K_2$) fixed
and search for the values of the frequency ratio presenting
localization. The result is shown in Fig.~\ref{fig:LocSpectrum}.
One clearly sees peaks at the simple rational values of
$r=f_2/f_1$. Each peak is associated with an increased number of
zero-velocity atoms, that is, an increased 
degree of localization.
The most prominent peaks are associated with integer values of $r$,
a rather natural result. Smaller peaks are associated with half-integers
values of $r$, even smaller ones with $r=p/3$ rational numbers, etc...
All these features are very well reproduced by the numerical
simulation (performed as described above, with no adjustable parameter)
shown in the inset of Fig. \ref{fig:LocSpectrum}.
The fact that the simulation displays
exactly the same behavior proves that it is not due
to an experimental artifact. 
Classical numerical calculations performed with the same
parameters do not show any kind of localization, neither in the
rational nor in the irrational case. The peaks are thus a purely
quantum feature.

We have also checked that the observed behavior does not
sensitively change when $f_1$ is varied. This rules out the possible
role of the so-called quantum resonances where the dynamics
is dominated by the quasi-degeneracy between unperturbed Floquet 
eigenstates. 
The observed width of the 1:1 resonance
is about 300 Hz, in
good agreement with the numerical calculation. A detailed
study of its width will be presented in the near future.

In conclusion, we have shown that, in the presence of a quasi-periodic
driving with two base frequencies, 
the kicked rotator does not show any ``short time"
dynamical localization
except when the ratio of the frequencies is close to a rational
number. In the latter case,
the system is time-periodic and displays clear evidence of dynamical
localization.
This conclusion is drawn from experiments performed with both
50 and 100 primary kicks, whereas the localization time is
of the order of 15 kicks. Longer kick sequences are impossible
because of the free fall of the atoms under
gravity, but numerical simulations show the same behavior
up to few thousands kicks. Although it is currently impossible,
experimentally or numerically, to decide if the DL is effectively
suppressed by the secondary kicks or if it corresponds to
a much longer localization time, the results presented here
clearly evidence a dramatic change in the behavior of the system
due to a secondary irrational frequency.
Furthermore, the destruction of DL by a secondary frequency 
is found to be a very sensitive phenomenon.

The authors are grateful to M. Druon, J. Pesez and J. Courbe 
for aid with the
experiments. Laboratoire de Physique des Lasers, Atomes et
Mol{\'e}cules (PhLAM)
is UMR 8523 du CNRS et de l'Universit{\'e} des Sciences
et Technologies de Lille. Centre d'Etudes et Recherches
Lasers et Applications (CERLA) is supported by Minist\`{e}re de la
Recherche, R\'{e}gion Nord-Pas de Calais and Fonds Europ\'{e}en de
D\'{e}veloppement Economique des R\'{e}gions (FEDER).
Laboratoire Kastler Brossel de
l'Universit\'e Pierre
et Marie Curie et de l'Ecole Normale Sup\'erieure is
UMR 8552 du CNRS.
CPU time on Cray C98 and T3E computers has been provided by IDRIS.

\vspace{1cm}

\vspace{0.5cm}

\vspace{1cm}

\end{document}